\documentclass[a4paper,twocolumn,superscriptaddress,11pt,unpublished]{quantumarticle}
\pdfoutput=1
\usepackage[utf8]{inputenc}
\usepackage[english]{babel}
\usepackage[T1]{fontenc}
\usepackage{amsmath}
\usepackage{amssymb}
\usepackage{hyperref}
\usepackage[numbers,sort&compress]{natbib}
\usepackage{braket}
\usepackage{tikz}
\usepackage{lipsum}
\usepackage{makecell}
\PassOptionsToPackage{normalem}{ulem}
\usepackage{ulem}

\def\ocq{\ket{0}\bra{0}\otimes I_N}
\DeclareMathOperator{\tr}{Tr}

\begin{document}

\title{Benchmarking quantum processors with a single qubit}
\date{\today}
\author{Oktay G\"okta\c{s}}
\affiliation{Agnostiq Labs, 100 College St, Toronto, Ontario, M5G 1L5, Canada}
\affiliation{Department of Physics and Center for Quantum Information and Quantum Control, University of Toronto, 60 St George St, Toronto, Ontario, M5S 1A7, Canada}
\email{Oktay@agnostiqlabs.com}
\author{W.K. Tham}
\email{wtham@physics.utoronto.ca }
\affiliation{Department of Physics and Center for Quantum Information and Quantum Control, University of Toronto, 60 St George St, Toronto, Ontario, M5S 1A7, Canada}
\author{Kent Bonsma-Fisher}
\email{kent.bonsma-fisher@nrc.ca}
\affiliation{Department of Physics and Center for Quantum Information and Quantum Control, University of Toronto, 60 St George St, Toronto, Ontario, M5S 1A7, Canada}
\affiliation{National Research Council of Canada, 100 Sussex Dr., Ottawa, Ontario, K1A 0R6, Canada}
\author{Aharon Brodutch}
\email{brodutch@physics.utoronto.ca}
\affiliation{Department of Physics and Center for Quantum Information and Quantum Control, University of Toronto, 60 St George St, Toronto, Ontario, M5S 1A7, Canada}
\affiliation{The Edward S. Rogers Department of Electrical and Computer Engineering, University of Toronto, 10 King’s College Road, Toronto, Ontario M5S 3G4, Canada}
\orcid{0000-0001-5536-1485}

\maketitle

\begin{abstract}
The first generation of small noisy quantum processors have recently become available to non-specialists who are not required to understand specifics of the physical platforms and, in particular, the types and sources of noise.
As such, it is useful to benchmark the performance of such computers against specific tasks that may be of interest to users, ideally keeping both the circuit depth and width as free parameters. 
Here we benchmark the IBM Quantum Experience using the  Deterministic Quantum Computing with 1 qubit (DQC1) algorithm originally proposed by Knill and Laflamme in the context of liquid state NMR. 
In the first set of experiments we use DQC1 as a trace estimation algorithm to produce visibility plots. 
In the second set we use this trace estimation algorithm to  distinguish between knots, a classically difficult task which is known to be complete for DQC1. 
Our results indicate that the main limiting factor is the length of the circuit, and that both random and systematic errors become an issue when the gate count increases. 
Surprisingly, we find that at the same gate count wider circuits perform better, probably due to randomization of coherent errors. 
\end{abstract}

\section{Introduction}

Small noisy quantum processors can now be implemented in various platforms and architectures including superconducting circuits \cite{IBMQ,Takita2017,Pokharel2018}, trapped ions \cite{Figgatt2017}, optics \cite{Rudolph16} and NMR \cite{Lu2017a}. These and other near-future processors are not expected to be universal for quantum computation \cite{Preskill2018quantumcomputingin} and need to be benchmarked in tasks that are suitable for noisy processors with little or no error correction.  The deterministic quantum computing with one qubit (DQC1) algorithm, which was originally developed for noisy NMR quantum processors, offers a good way to benchmark these processors. In this work we benchmark two IBM quantum processors, first using simple DQC1 circuits to calculate the trace of a unitary, and then, in a specific task, using DQC1 to distinguish between knots. 

The experiments used between 3 and 8 qubits and were initially run on the IBM Q 16 Rüschlikon \cite{IBM-r} and later on the IBM Q 14 Melbourne\cite{IBM-m}. 
The first set of experiments (Sec. \ref{sec:DQC1bench}) involved the estimation of the normalized trace of  1 and 3 qubit unitaries. The results allow us to make some general statements about the noise in the circuit, in particular depolarizing noise and systematic (coherent) errors.
Somewhat surprisingly, the performance of the 3 qubit algorithms as a function of the number of gates was better than the 1 qubit algorithms, most likely due to the reduction in correlated noise when the gates act on different qubits. In the second set of experiments, we used the DQC1 algorithm to evaluate various Jones polynomials (Sec. \ref{sec:jones}).  
The results show that while the evaluated Jones polynomials
tend to be far from theoretical values, the errors are consistent for the different circuits. 
This implies that the processors can be used to distinguish between various knots made by closing a braid of up to 3 strands, as long as the evaluations are run at approximately the same time (i.e., not days apart) using the same subset of qubits, such that systematic errors in gate operations remain approximately the same from run to run. 

\section{DQC1}
The DQC1 model was originally proposed by Knill and Laflamme \cite{Knill1998a} in the context of room temperature, liquid state NMR quantum computing where the initial (thermal)  state $\rho_i$ is very noisy. 
As a consequence of the noise, the signal-to-noise ratios in the readout are small and the computation is done on an ensemble with ensemble readout, i.e., the result is an estimate of the expectation value of some observable.  
Knill and Laflamme noted that in an $N+1$ qubit NMR processor it is possible to prepare an initial state of the type $\rho_i=[\alpha\ket{0}\bra{0}+(1-\alpha)I_1]\otimes I_N$ (where $I_n=\frac{1}{2^n}\openone_n$ is the $n$ qubit maximally mixed state) efficiently\footnote{The initialization procedure works for some fixed $\alpha \ll 1$ that depends on the parameters of the experiment, and does not scale badly with $N$.}.  
Under the assumption that the evolution is given by a unitary operator $V$, the final readout on the first qubit will be $\tr (V \rho_i V^\dagger\sigma_k^{(1)})=\alpha\tr (V \left ( \ocq \right ) V^\dagger\sigma_k^{(1)})$ where $\sigma_k^{(1)}$ is a Pauli operator on the first qubit and  $k\in\{x,y,z\}$. 
Noting that the polarization  parameter $\alpha$ is simply used to re-scale the expectation value, it is possible to assume $\alpha=1$ without loss of generality, as we will do throughout this work. 
Under this assumption the first qubit is initially pure, or ``clean'', while the other qubits are completely mixed. This model is therefore sometimes called the `one clean qubit' model~\cite{Shor2007}.


\begin{figure}[h]
  \centering
\includegraphics[width=0.7\columnwidth]{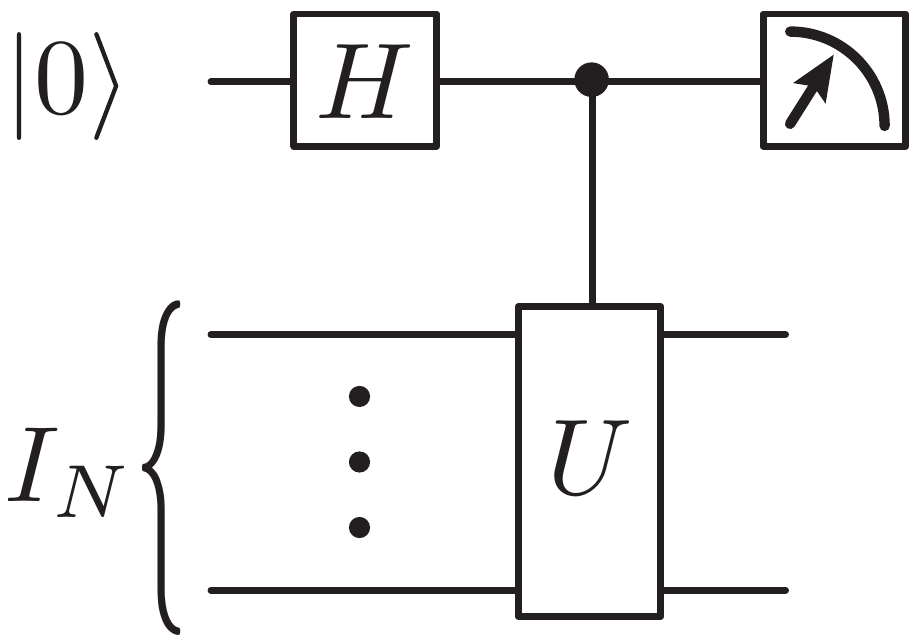}
  \caption{A DQC1 circuit for estimating the trace of a unitary $U$ with the first qubit initialized in a pure state. The same computation is run a large number of times with the final measurement is cycled between $\sigma_x$ and $\sigma_y$ to get an estimate of the the real and imaginary parts respectively.}
  \label{fig:DQC1}
\end{figure}

In the DQC1 model the classical input describes the unitary operator $V$ which is assumed to have an efficient description, i.e., it can be decomposed into a (polynomial in $N$) sequence of one and two qubit gates.
It is common to further restrict $V$ to a Hadamard operator on the first qubit, followed by a controlled unitary from the first qubit, $U$, targeting all other qubits, i.e., $V=\left(\ket{0}\bra{0}\openone_N+\ket{1}\bra{1} U\right)H^{(1)}$. 
Here $U$ is an $N$ qubit unitary with an efficient classical description and $H^{(1)}$ is a Hadamard operator on the first qubit (this is sometimes called cDQC1~\cite{Boyer2017}). 
In this restricted model, the state of the first qubit at the end of the computation (before readout) is given by 
\begin{equation}
\rho^{(1)}_f=\frac{1}{2}\left(I_{1}+\frac{\text{Tr}U}{2^{N}}\left|1\right>\left<0\right|+\frac{\text{Tr}U^{\dagger}}{2^{N}}\left|0\right>\left<1\right|\right)
\end{equation}
so that $\frac{1}{2^N}\tr U=\langle \rho_f^{(1)}(\sigma_x+i\sigma_y)\rangle$. 
That is, the model can be used to estimate the normalized trace of the unitary $U$. 
A number of results suggest that the trace estimation algorithm cannot be simulated efficiently by a classical computer, with some recent results including the use of DQC1 for parity learning \cite{Park2018}, and a sampling version \cite{Morimae2017}  which follows the standard definition above, but allows single shot readout.

Shor and Jordan~\cite{Shor2007} used the DQC1 model to define a computational complexity class. They then showed that the trace estimation algorithm is computationally equivalent to the full DQC1 model and furthermore showed that adding a small (at most logarithmic in $N$) number of pure qubits does not change the computational power of the model. They also showed that the estimation of the Jones polynomial for the trace closure of a braid at the fifth root of unity (a problem in  knot theory, see Sec. \ref{sec:jones} below) is DQC1 complete.

\subsection{Noise in  DQC1}

The DQC1 model is designed to handle noisy initial states but, to the best of our knowledge, its performance under noisy dynamics has not been analyzed.  
An $N$ qubit unitary $V$ can generally be decomposed into a sequence of fundamental unitaries $\{W_k\}$ such that $V=\prod_k W_k$. Ideally, these fundamental unitaries correspond to gates that are physically implementable on the processor. But in practice, the gates are imperfect and errors that are often difficult to characterize degrade the computation\cite{Emerson2007a,Wallman2015}.

One fairly simple model is to assume depolarizing noise, where each gate is a probabilistic mixture of the desired unitary $W_k$ and a completely depolarizing channel. 
The ideal transformation $\rho\rightarrow W_k\rho W_k^\dagger$ of the state is replaced by $\rho \rightarrow \alpha_k W_k \rho W_k^\dagger+(1-\alpha_k)\openone$ where $\alpha_k$ is the purity of the channel. All subsequent unitary operations and depolarizing channels leave the identity unchanged, so the full sequence  will be $\alpha V\rho_iV^\dagger+(1-\alpha)\openone$  with $\alpha=\prod_k \alpha_k$. 

The fact that the purity falls exponentially with the number of gates does not bode well for the computation.  Standard quantum error correction methods rely on a supply of pure qubits so they are not suitable for DQC1. However, for our purposes, focusing on small or intermediate size processors, the issue of exponential noise might not be debilitating especially when one considers a logical implementation of the model using some clean physical qubits and single shot measurements to combat errors.

The relatively simple behavior of the DQC1 algorithm in the presence of depolarizing noise makes it a good tool for benchmarking against a depolarizing noise model. In our results below, we used the $R^2$ of a fit to the depolarizing noise model as a benchmark. The behavior of the circuits as a function of the number of gates provided evidence that the most significant source of error was a systematic error in the CNOT gates. 

\section{Implementation of the algorithm}

\begin{figure}[h]
  \centering
\includegraphics[width=\columnwidth]{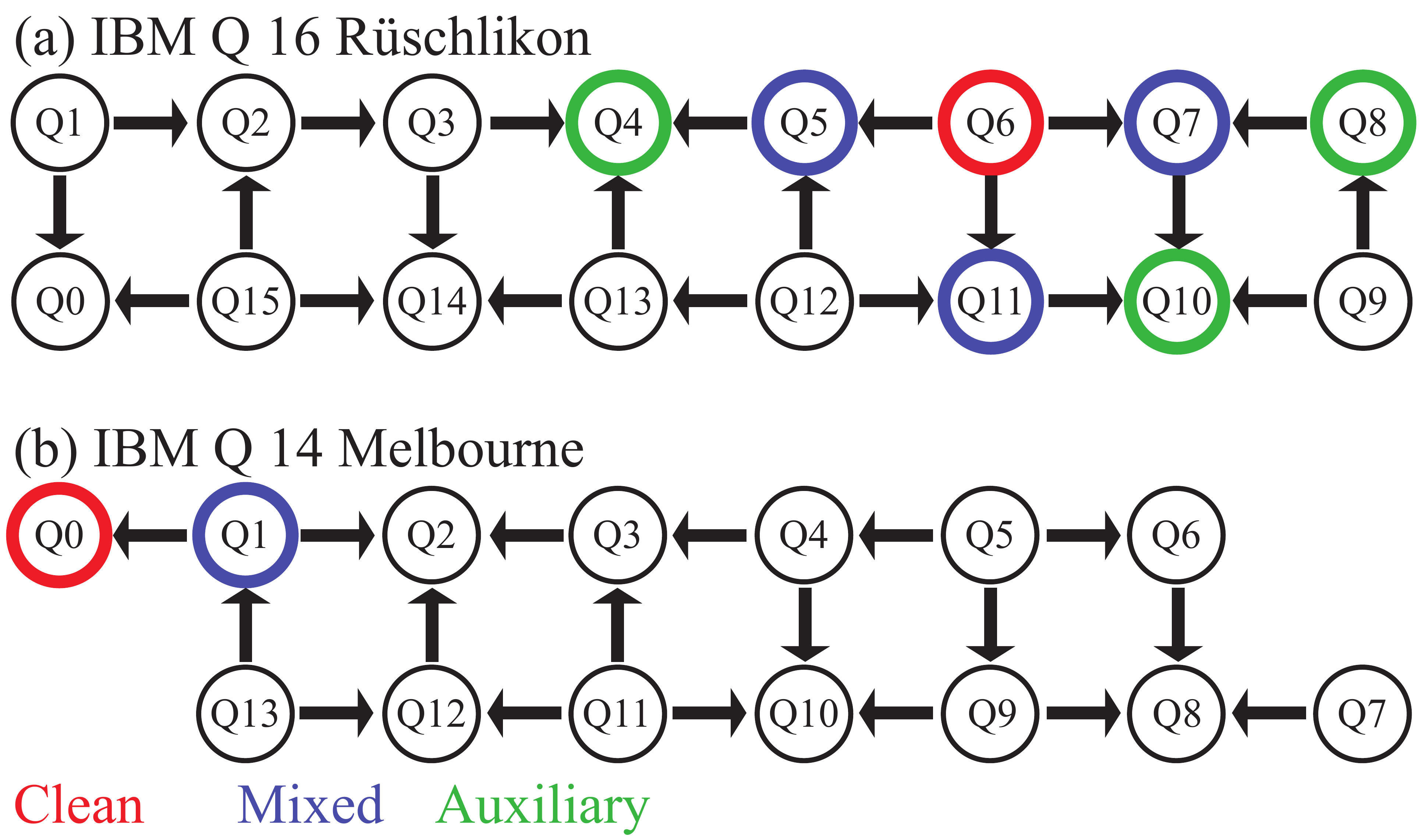}
  \caption{(a) Qubits used on the IBM Q 16 R{\" u}schlikon chip. Qubit 6 (red) was the clean qubit. Qubits 5, 7, and 11 (blue) were the mixed qubits. Mixed states were generated by first performing an entangling operation with auxiliary qubits 4, 8, and 10 (green), respectively. Black arrows show the control-target relationship for coupled qubits.  The statistical uncertainty is upper bounded by $1/2^{7.5}$. (b) The IBM Q 14 Melbourne was used for the knot experiments. Qubits 0 and 1 were used for the pure and mixed states, respectively, in the first set of knot experiments. Subsequent experiments used all 18 pairs of connections between qubits.  }
  \label{fig:IBM16}
\end{figure}

We implemented DQC1 on IBM superconducting qubit quantum processing units (QPU) via a web-based application programming interface (API). The code is available online \cite{Code} and a technical descriptions of the processors can be found in Ref.~\cite{IBMQ}. All results described in the present work are limited to data obtained via the web API, and not direct physical access to IBM hardware. Basic tests to benchmark DQC1 performance on gate-based machines, described in Section~\ref{sec:DQC1bench}, were executed on the 16-qubit ``R{\" u}schlikon'' QPU. Application of DQC1 to a useful task (evaluation of Jones polynomials), described in Section~\ref{sec:jones}, was executed on the 14-qubit ``Melbourne'' QPU.

A single qubit (shown in {\color{red}{red}} in Fig.~\ref{fig:IBM16}) was designated the ``clean'' qubit whereas another disjoint subset of qubits (shown in {\color{blue}{blue}} in Fig.~\ref{fig:IBM16}) was chosen to be ``noisy''. 
A gate-based QPU, however, is usually designed to operate with pure states under unitary evolution as much as possible. To prepare these ``noisy'' qubits in the $\left(\mathbb{I}/2\right)^{\otimes N}$ state, we used two techniques. In the first set of experiments we first entangled each qubit with an adjacent qubit to produce the (pure) Bell state $\left|\Phi^{+}\right>= (\left|00\right>+\left|11\right>)/\sqrt{2}$, and then ignored (or traced away) that adjacent qubit. This approach introduces a 2X overhead in the number of qubits required for state preparation of these ``noisy'' qubits.
For the estimation of the Jones polynomial, we used bit flip on the ``noisy'' qubit for half the experiments and averaged over the results. This method can be modified for a multiple qubit scenario by randomly flipping all ``noisy'' qubits and averaging over the results.

\section{Trace estimation on the quantum processor}\label{sec:DQC1bench}

\begin{figure*}
  \centering 
 \includegraphics[width=0.99\textwidth]{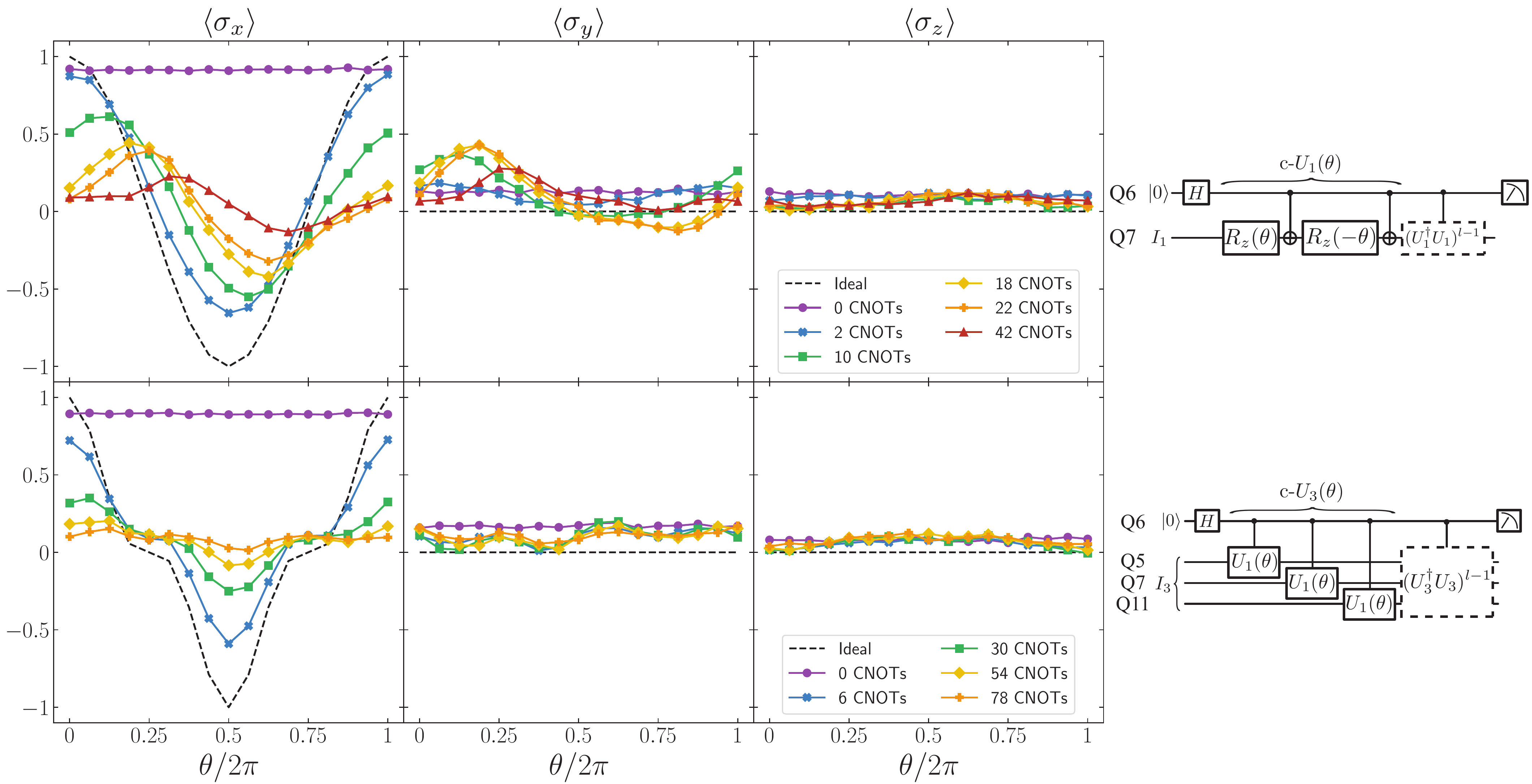}
  \caption{Expectation values $\left \langle \sigma_x \right \rangle$, $\left \langle \sigma_y \right \rangle$ and $\left \langle \sigma_z \right \rangle$ when applying the c$-U(\theta)$ from the control qubit to one mixed qubit (Q7, top row), and three mixed qubits (Q5, Q7, Q11, bottom row). Colored curves represent applying the c$-U(\theta)$ multiple times(The data for 0 CNOTs means no operation applied at all). As expected, the deviations from the theoretical curve get larger as the number of gates increases, however the gate errors are not random and the contribution of non-depolarizing noise is significant. This is especially apparent in the $\sigma_y$ plot which is expected to be near 0 at all times. Note that for large numbers of CNOT gates the deviation from 0 is far above the statistical uncertainty which is upper bounded by $1/2^{7.5}<0.01$. The coherent errors  are partially suppressed in the 1+3 qubit circuit, probably due to coherent errors being averaged out over the different qubits.}
  \label{fig:res1} 
\end{figure*}

We implemented the $N=1$ version of the trace estimation algorithm (Fig.~\ref{fig:DQC1}) with \[U_{N=1}^{(l)}(\theta)=U_1(\theta)(U_1(\theta)^{\dagger} U_1(\theta))^{l-1}\] where $l\ge 1$ is the number of repetitions  and
$U_{1}(\theta)=e^{-i\theta/2} \left|0\right>\left<0\right| + e^{i\theta/2}\left|1\right>\left<1\right|$ (see top RHS of Fig. \ref{fig:res1}). 
We also implemented the $N=3$ version, replacing $U_{1}^{(l)}(\theta)$ with $U_{3}^{(l)}(\theta)=U_{1}^{(l)}(\theta)^{\otimes3}$ so \[U_{N=3}^{(l)}(\theta)=U_3(\theta)(U_3(\theta)^{\dagger} U_3(\theta))^{l-1}\] (see bottom RHS of Fig. \ref{fig:res1}).

The circuit was chosen to maximize contrast with respect to $\theta$  (i.e., $\tr[U_{1}^{(l)}(0)]=1$ and $\tr[U_{1}^{(l)}(\pi)]=-1$) for \emph{any} value of $l$. Increasing $l$ merely introduces repetition of a gate sequence that should, logically, be equivalent to the identity. In practice, however, gate errors and noise means increasing $l$ yields noisier outputs. Results for a final measurement of $\sigma_x$, $\sigma_y$ and $\sigma_z$ are shown in Fig. \ref{fig:res1}.

\begin{figure}[h]
  \centering 
   \includegraphics[width=0.75\columnwidth]{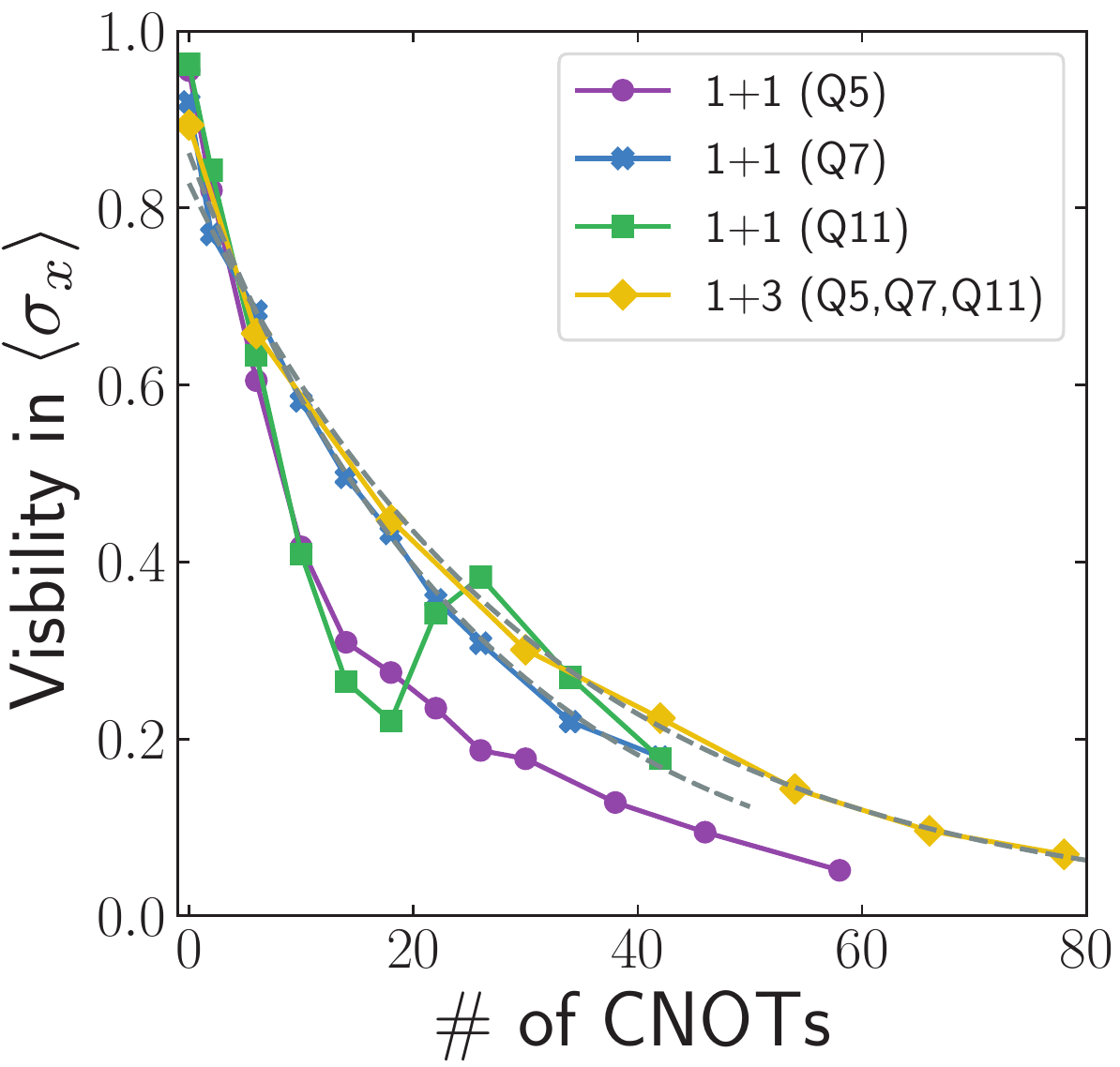}
  \caption{Visibility in $\left \langle \sigma_x \right \rangle$ as a function of circuit length. Decay in visibility is indicative of noise in the system. Assuming purely depolarizing noise decay is exponential. Grey dashed lines are fits of $f(x) = a e^{-x/\tau}$ to the 1+1 (Q7) and 1+3 (Q5,Q7,Q11) data.}
  \label{fig:visplot}
\end{figure}

As expected, the results deviate further from the ideal as we go to higher gate counts. The reduction of the absolute values in the $\sigma_x$ plots can be attributed to depolarizing noise, however the fact that the shape changes in all three plots (and in particular the deviations from 0 in $\sigma_y$), indicate a coherent error, probably as a result of a systematic error in the CNOT gates. 
For a quantitative indication of the coherent errors (more precisely the deviation from depolarizing noise) we  defined the visibility
\[\text{Vis}=\max_{\theta}\left\langle \sigma_x \right\rangle.\] 
This function decays exponentially with the number of gates when imperfections are due to depolarizing noise. A plot of visibility as a function of the number of gates  (Fig. \ref{fig:visplot}) shows that this is clearly not the case for qubit 11 (paired with 5) where there is a spike in visibility around 20 CNOT gates. The other couplings show the expected qualitative behaviour, but a fit to a depolarizing noise model shows some deviations (see Table~\ref{tab:noise}). 

Generally, it is possible to convert biased noise channels into a depolarizing channels by adding some randomness and averaging. This is apparent when comparing the 1+1 and 1+3 qubit results.  In Table \ref{tab:noise} we see that  fit for the 1+3 qubit result is better than each of the individual results. This is most likely a result of averaging over different coherent errors for the 3 different pairs of interacting qubits. We note that in general this is not an indication of better performance overall. Though the exponential depolarizing rate appears slower in the 1+3 qubit case, the circuit will have to be 3 times longer to perform a similar task.  

\begin{table}[h]
\begin{tabular}{c|c|c}
Qubits & $\tau$ & $R^2 = 1 - \frac{\sum_i (y_i - f_i)^2}{\sum_i (y_i - \bar y)^2}$
\\ \hline 
Q5 & 24.50 & 0.933
\\
Q7 & 25.81 & 0.995
\\
Q11 & 28.91 & 0.695
\\
Q5, Q7, Q11 & 30.75 & 0.997 
\end{tabular}
\caption{Assuming only depolarizing noise, we fit a decay model $f(x) = a e^{-x/\tau}$ to the data in Fig.~\ref{fig:visplot} (linear fit to the logarithm of the data). }\label{tab:noise}
\end{table}

\begin{figure}[h!]
  \centering 
 \includegraphics[width=0.75\columnwidth]{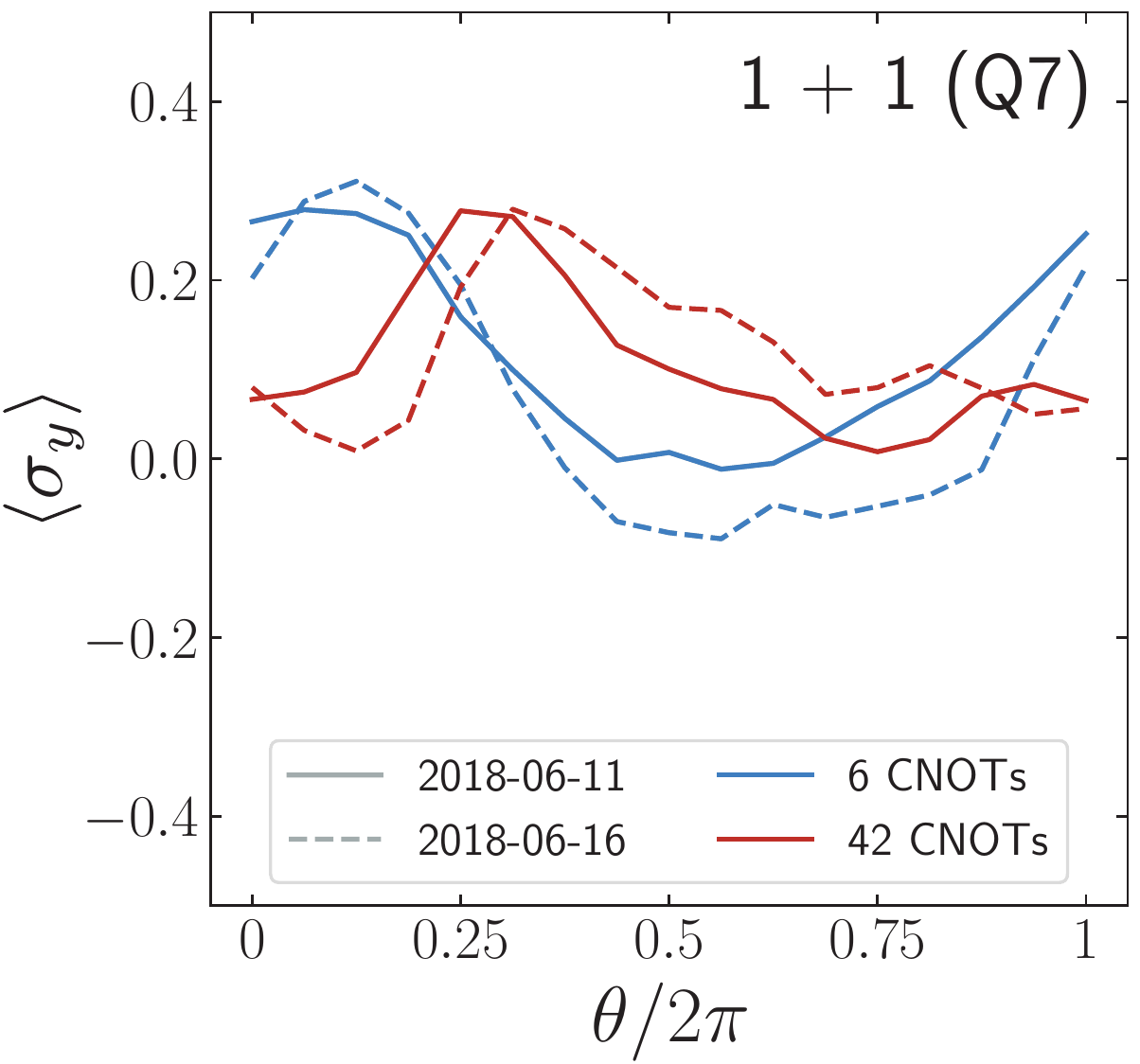}
  \caption{$\sigma_y$ expectation value of 1+1 data (qubit 7) taken on different dates, 5 days apart. The variations between results at different times, is far greater than the statistical uncertainty which is upper bounded by $1/2^{7.5}<0.01$, and indication that systematic calibration errors change significantly over time. }
  \label{fig:diffdays}
\end{figure}

Running the experiment at different times produced different results, in particular the systematic (coherent) errors were not consistent over long periods of time.  Results for the expectation value of $\sigma_y$ taken on the same pair of qubits 5 days apart are plotted in Fig.~\ref{fig:diffdays}. Since the theoretical expectation value should be constant ($\langle \sigma_y \rangle =0$) the plots are a good indication of coherent errors which, even at a circuit depth of 6 CNOT gates, produce a visibly different plot.

\section{Distinguishing knots with Jones polynomials}\label{sec:jones}

\begin{figure}
  \centering
\includegraphics[width=\columnwidth]{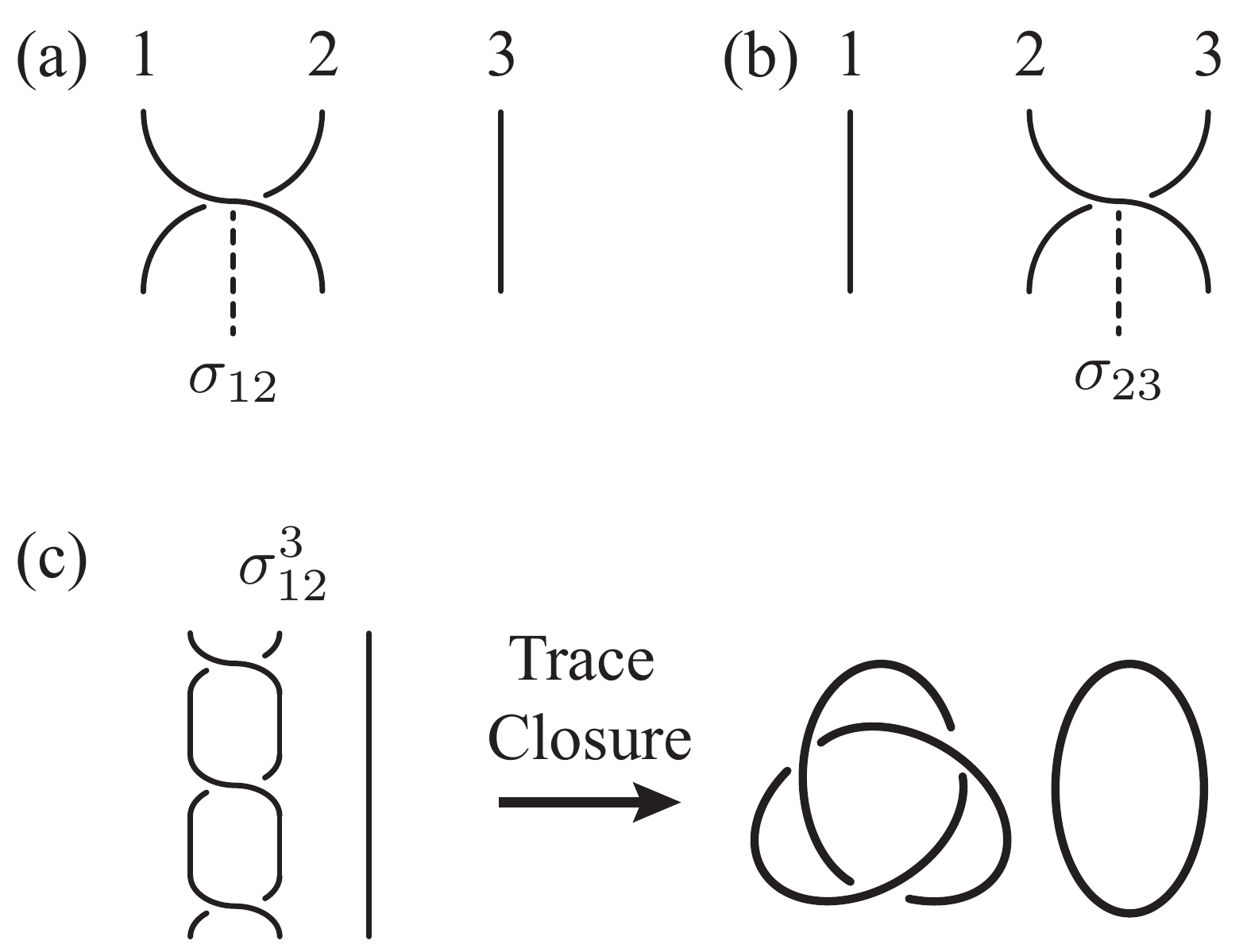}
  \caption{Visualization of $\sigma_{12}$ (a) and $\sigma_{23}$ (b) crossing operations on three strands. (c) The braid word $\sigma_{12}^3$, three consecutive $\sigma_{12}$ crossings. By taking the trace closure, connecting the bottom of a strand to its top, the first two strands form the trefoil knot while the third strand forms the unknot. The braid closures of (a) and (b) are topologically equivalent, but the different braid words lead to different circuit implementations in the experiment.  }
  \label{fig:knots}
\end{figure}

The task of identifying whether two knots (smooth closed curves in $\mathbb{R}^3$) are topologically equivalent has implications beyond mathematics, reaching into statistical mechanics, quantum field theory and quantum gravity~\cite{baez1994gauge, pullin1994}.  
Knots can be faithfully represented in two-dimensional pictures, and so the task can be recast as determining whether two pictures of knots can be made equal using transformations called Reidemeister moves.  
This task is computationally intensive and even in  the simplest case, identifying a knot as the \emph{unknot}, there is no known efficient solution \cite{Lackenby2015}.
Here we consider a particular type of knot, the trace closure of a 3-strand braid, which  can be represented by drawing a two-dimensional braid and closing each end at the bottom with the associated strand at the top (right-most to right-most etc., see Fig.~\ref{fig:knots}(c). Another type of knot, the plat closure, is constructed by connecting adjacent strand ends at the top and at the bottom. 

The Jones polynomial is a complex function invariant for oriented knots~\cite{Jones1985, Jones1987}, which  allows one to distinguish one knot from another. 
However, constructing the polynomial for a given two-dimensional picture of a knot is not trivial. The number of terms in the polynomial (before simplification) scales exponentially with the number of strand crossings.
Evaluating the Jones polynomial at a single point would give sufficient information to tell if two knots are different. Equivalent knots must have the same Jones polynomial value at any given point, whereas knots that are not the same might not. 
Approximating the value of the Jones polynomial at $e^{2\pi i/5}$ is a particularly interesting task for quantum computers. If one takes the plat closure of a braid, rather than the trace closure, the task is known to be BQP-complete~\cite{Aharonov2009}.

In 2007 Shor and Jordan~\cite{Shor2007} showed that approximating the polynomial at $e^{2\pi i/5}$ for the \emph{trace} closure of a braid is a complete task for DQC1. Passante \emph{et al.}~\cite{Passante2009} demonstrated this task in a four-qubit liquid-state NMR processor, studying knots with four strands and multiple crossings.  Here we use DQC1 to approximate the Jones polynomial for knots of three strands for the purpose of benchmarking the IBM Q 14 Melbourne quantum processor. We consider knots which are constructed by taking the trace closure of braid words up to 9 crossings (see Fig.~\ref{fig:knots}). We study the same knots constructed by multiple iterations of either the $\sigma_{12}$ crossing (first strand over the second), or the $\sigma_{23}$ crossing (second over the third). 
Since approximating the values of Jones polynomials with a noisy machine is difficult, we are content with the ability to classify knots as different when they are indeed different. As in the previous sections, we do not perform any type of error mitigation during the computation or in post-processing, apart from simplifying the circuit to require fewer gates. 

Following the treatment in Ref.~\cite{Passante2009}, the unitary used in the DQC1 protocol is related to the braid through the Fibonacci representation. For a three-strand braid, the $\sigma_{12}$ and $\sigma_{23}$ unitaries are given by
\begin{equation}
\sigma_{12} = 
\begin{pmatrix}
    a & 0 & 0 & 0 \\
    0 & b & 0 & 0 \\
    0 & 0 & a & 0 \\
    0 & 0 & 0 & 1
\end{pmatrix}
, \quad
\sigma_{23} = 
\begin{pmatrix}
    e & d & 0 & 0 \\
    d & c & 0 & 0 \\
    0 & 0 & a & 0 \\
    0 & 0 & 0 & 1
\end{pmatrix},
\end{equation}
where $a = e^{3\pi i /5}$, $b = e^{-4\pi i /5}$, $c = \frac{b}{\phi^2} + \frac{a}{\phi}$, $d = \frac{b-a}{\phi^{3/2}}$, $e = \frac{b}{\phi} + \frac{a}{\phi^2}$, and $\phi  = (1 + \sqrt{5})/2$. For braids with $n$ strands we require $m \times m$ sized unitaries where $m$ is the $n^\text{th}$ number in the Fibonacci sequence. For 3 strands the unitaries map between 3 states, requiring 2 qubits. This results in an unused portion of the Hilbert space -- the $\ket{11}$ state is not used for the approximation and adds a constant term to the trace which is straightforwardly dealt with.

\begin{figure}[h]
  \centering 
  \includegraphics[width=\columnwidth]{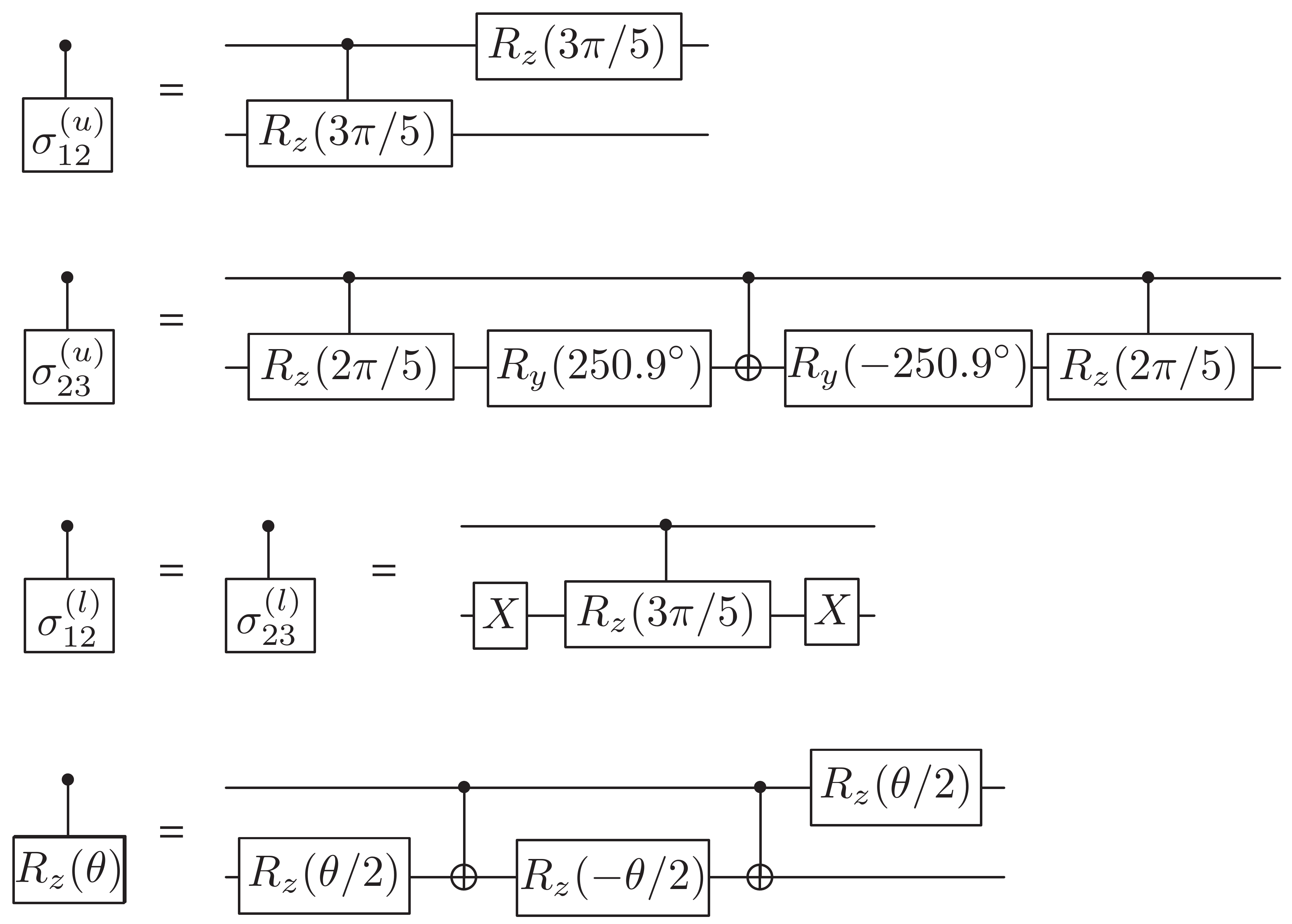}
  \caption{
  Circuits for the controlled $\sigma_{12}^{(u)}$
  , $\sigma_{23}^{(u)}$, $\sigma_{12}^{(l)}$ and $\sigma_{23}^{(l)}$ unitaries.
  Here, $R_{y}\left(\theta\right)=\left(\protect\begin{array}{cc} \cos\theta & \sin\theta\protect\\ -\sin\theta & \cos\theta \protect\end{array}\right)$,
  and $R_{z}\left(\theta\right)=\left(\protect\begin{array}{cc} 1 & 0\protect\\ 0 & e^{i\theta} \protect\end{array}\right)$.
  }
  \label{fig:knotcircuit}
\end{figure}


To implement DQC1 we must construct controlled versions of the braid word unitaries. However, a single controlled-$\sigma_{23}$ operation requires approximately 50 CNOT gates, meaning any braid word would be prohibitively long on current quantum processors. Indeed, Sec.~\ref{sec:DQC1bench} shows that $\left \langle \sigma_x \right \rangle$ and $\left \langle \sigma_y \right \rangle$ on the clean qubit would decay substantially after just one controlled-$\sigma_{23}$ gate. To simplify the problem we use the fact that both $\sigma_{12}$ and $\sigma_{23}$ are block diagonal, and so any braid word will also be. We perform a controlled version of each block of a braid word in separate experiments, measure their traces via the clean qubit, and combine them afterwards. The controlled implementations of the blocks require substantially fewer CNOT gates:  $\sigma_{12}^{(u)}$, $\sigma_{23}^{(u)}$, and $\sigma_{12}^{(l)} = \sigma_{23}^{(l)}$, where $u$ $(l)$ refers to the upper (lower) block of the unitary, can be performed with 2, 5 and 2 CNOT gates respectively.  The braidword for the trefoil knot outlined in Fig.~\ref{fig:knots}, $\sigma_{12}^3$, requires 6 CNOT gates for each of the upper and lower blocks. Performing the braidword $\sigma_{23}^3$, which gives the same knot upon taking the trace closure, requires 15 CNOT gates for the upper block and 6 for the lower. 

Having used the DQC1 protocol to measure the trace of each block of a braid word unitary, we then estimate the value of Jones polynomial at the fifth root of unity for each knot. To do this we combine the measurements on each block of the unitary, e.g., $U = \sigma_{12}^3$, to find the weighted trace of the braid word. First, we subtract off the contribution to the trace of the lower block, $U^{(l)}$, from the $\ket{11}$ state. We then add the traces of the two blocks together while weighting the upper block by a factor of $\phi$~\cite{Passante2009,Passante2012a},
\begin{eqnarray}
\text{WTr}\,U &=& \phi \times \tr U^{(u)} + \tr U^{(l)} - 1 \\
&=& \phi \times (\left \langle \sigma_x \right \rangle^{(u)} + i \left \langle \sigma_y \right \rangle^{(u)}) 
\nonumber \\
&& \quad \quad 
+ \left \langle \sigma_x \right \rangle^{(l)} + i \left \langle \sigma_y \right \rangle^{(l)} - 1.
\nonumber
\end{eqnarray}
We then calculate the Jones polynomial value as
\begin{equation}
    V_U(t= e^{2\pi i /5}) = (-(e^{2\pi i /5})^4)^{3w} \times \frac{1}{\phi} \text{WTr}\,U,
\end{equation}
where $w$ is the writhe of the knot, defined as the difference between left-over-right crossings ($\sigma_{12}$ and $\sigma_{23}$) and right-over-left crossings ($\sigma^\dagger_{12}$ and $\sigma^\dagger_{23}$). In this work we only consider knots with $w>0$.

\begin{figure}
  \centering 
 \includegraphics[width=\columnwidth]{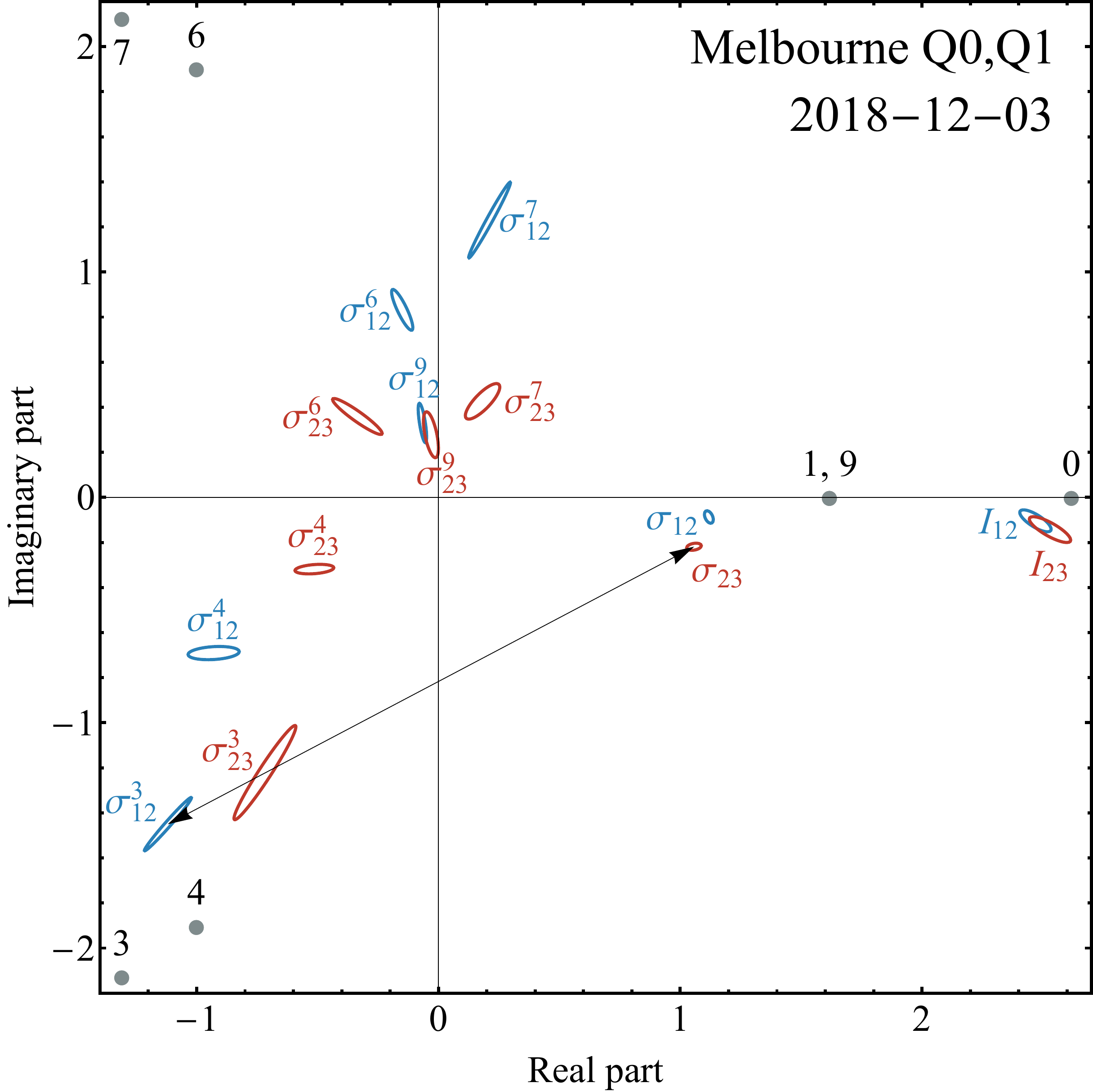}
  \caption{Results of Jones polynomial evaluations on qubits Q0 and Q1 of the IBM Q 14 Melbourne. Knots are given by the trace closure of  $\sigma_{12}^k$ (blue points), and $\sigma_{23}^k$ (red points), $k$ runs from 0 to 9 crossings. Gray points mark the theoretical Jones polynomial values for each knot, with numbers represent the crossings in the associated braid word, for each $k$ the knots represented by $\sigma_{12}^k$ and $\sigma_{23}^k$ are equivalent.  Ellipses represent the standard deviation of 12 trials ($2^{12}$ shots each) for each knot.  The distance between ellipses measures, in some sense, how well two knots can be distinguished on the IBM QPU. The black arrow between $\sigma_{12}^3$ and $\sigma_{23}$ marks the distance between the two Jones polynomial estimates with a similar gate count (see also Table~\ref{tab:JPdist}). For clarity, results from knots with 2, 5, and 8 crossings (which are closer to the center) are not plotted. Note that while the results generally get closer to the center as the gate count increases (a signature of depolarizing noise), results for 7 and 9 crossings appear in the wrong quadrant in both representations (a signature of systematic errors).  }
  \label{fig:JPresults}
\end{figure}

\begin{figure}
  \centering 
 \includegraphics[width=0.9\columnwidth]{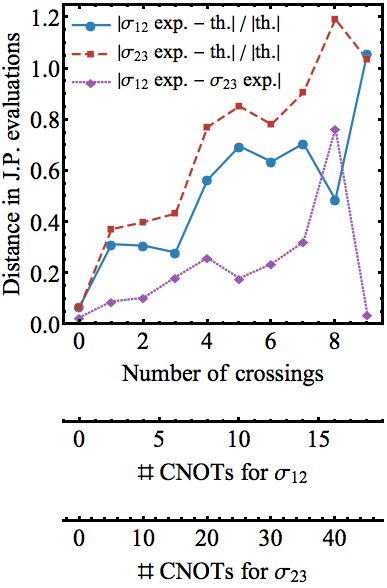}
  \caption{Distance between the evaluated Jones polynomial from $\sigma_{12}$ unitaries, $\sigma_{23}$ unitaries and theory for knots with varying numbers of crossings. Distances are normalized by the theoretical values of the polynomials to account for values close to the origin. $\sigma_{23}$ estimates deviate from the theoretical values for fewer crossings, likely due to the fact that a single $\sigma_{12}$ unitary requires 2 CNOTs while a $\sigma_{23}$ requires 5 CNOTs. The distance between polynomial estimates for $\sigma_{12}$ and $\sigma_{23}$ implementations increases with number of crossings (with the exception of the final point where they both tend the origin), making two versions of the same knot distinguishable from one another.
  }
  \label{fig:JPdist}
\end{figure}

In Fig.~\ref{fig:JPresults} we plot one set of results of the Jones polynomial, $V_U(e^{2\pi i /5})$, estimation for knots with 0 to 9 crossings, constructed solely by either $\sigma_{12}$ or $\sigma_{23}$ crossings. We find that as the numbers of crossings are increased the estimated polynomials quickly deviate from theoretical values (see Fig.~\ref{fig:JPdist}). This is to be expected from the studies in previous sections. As the circuit depth increases, the measured $\left \langle \sigma_x \right \rangle$ and $\left \langle \sigma_y \right \rangle$ on the clean qubit decay exponentially. This in turn causes the Jones polynomial values to tend to the origin on the complex plane with increasing circuit depths.  

Though the deviation from the theoretical value is not ideal performance, the principle behind estimating the Jones polynomial is to distinguish between knots. In this spirit, we note that the distance between the two implementations of each knot (using either $\sigma_{12}$ or $\sigma_{23}$) remains relatively low compared to the distance from their theoretical value. Differences between the two experimental implementations of the same knot are likely driven by the significantly higher circuit depth for each $\sigma_{23}$ unitary (5 CNOTs vs. 2 CNOTs).  Importantly, if we compare different knots that have the same circuit depth -- e.g., $\sigma_{12}^5$ and $\sigma_{23}^2$ each use 10 CNOT gate for their upper blocks, and represent different knots -- we see that they are largely distinguishable from one another when the gate count is low.

\begin{table}
\begin{tabular}{c|c|c|c}
\makecell{CNOTs \\ in $\sigma_{23}^k$} & Unitaries  & \makecell{J.P. dist. \\ (Exp.)} & \makecell{J.P. dist. \\ (Theory)}
\\ \hline 
5 & $| \sigma_{23} - \sigma_{12}^2 |$ & $1.42 \pm 0.05$ & 2.15
\\
5 & $| \sigma_{23} - \sigma_{12}^3 |$ & $2.50 \pm 0.08$ & 3.62
\\
10 & $| \sigma_{23}^2 - \sigma_{12}^5 |$ & $0.66 \pm 0.05$ & 1
\\
15 & $| \sigma_{23}^3 - \sigma_{12}^7 |$ & $2.6 \pm 0.2$ & 4.25
\\
15 & $| \sigma_{23}^3 - \sigma_{12}^8 |$ & $2.0 \pm 0.2$ & 3.24
\end{tabular}
\caption{Comparison of Jones polynomial estimates for implementations of different knots using the similar circuit depths. A large distance, relative to the error, indicates that two knots can be distinguished using the IBM QPU.   }\label{tab:JPdist}
\end{table}



\begin{figure}
  \centering 
 \includegraphics[width=\columnwidth]{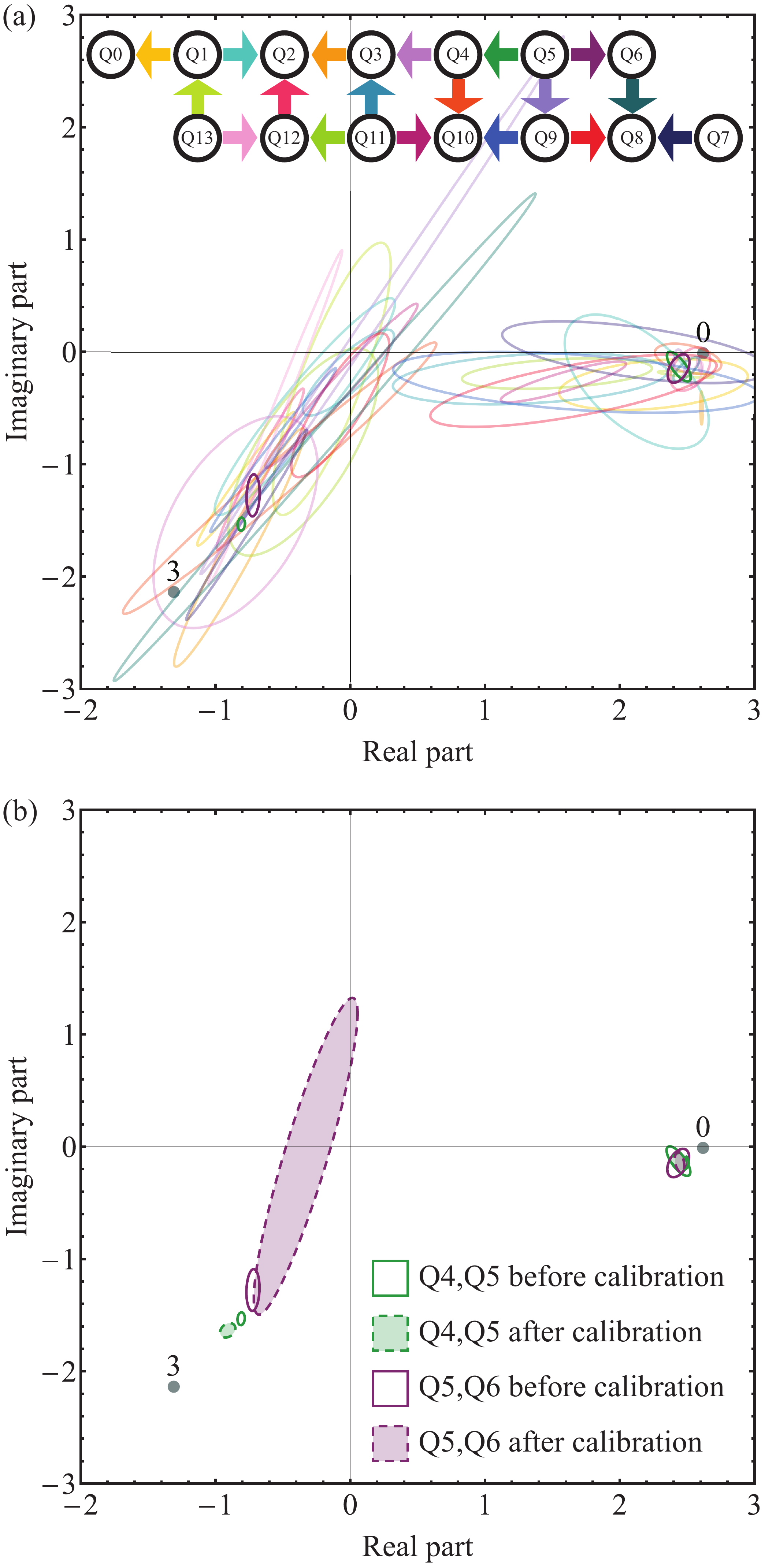}
  \caption{(a) Jones polynomial results of the $\sigma_{12}^0$ (right) $\sigma_{12}^3$ (left) unitaries using different qubit combinations on the Melbourne processor. 
  Shown in bold are the Q4-Q5 (dark green) and Q5-Q6 (dark purple) pairs which gave the best performance.
  Computations were performed on February 22, 2019 (before calibration). 
  (b) Comparing qubit pairs Q4-Q5 (dark green) and Q5-Q6 (dark purple) before (solid) and after (dashed, filled) the February 25, 2019 re-calibration. 
  The second set of computations was performed on February 26, 2019 (after calibration). A routine calibration process was performed on the machine by the IBM team between these two dates.
  For both (a) and (b) ellipses represent the standard deviation of estimating Jones polynomials over 10 trials (1024 shots each). 
  }
  \label{fig:Calibration}
\end{figure}

At higher gate counts the values are not only closer to the origin (as expected), but also behave qualitatively different from the theoretical results, for example in Fig. \ref{fig:JPdist} the real part of the braid with 7 crossings should be more negative than that of 6 crossings, but in both implementations ($\sigma_{12}$ and $\sigma_{23}$) the  knot with 7 crossings is positive while the knot with 6 is negative. Moreover these types of errors, while fairly consistent on a single run of the experiment, appear to be very different when the experiment is repeated later and/or on different qubits. 
In Fig.~\ref{fig:Calibration}(a) we show the results for 0 and 3 crossings ($\sigma_{12}$)  taken on all 18 different pairs of qubits. The results indicate that both the mean and the spread depend on the choice of qubits. In Fig.~\ref{fig:Calibration}(b) we compare the results of the braids with 0 and 3 crossings ($\sigma_{12}$) at different times, where we chose the qubits pairs Q5,Q6 and Q4,Q5 for best performance. Even at relatively low gate counts, we observe deviations from one run of the experiment to the next. 
One consequence of these results is that there is no simple way to correct for errors in post-processing. 
Such corrections would have been possible if the dominant source of error was depolarization, in which case we could multiply the results by a factor that depends on the number of gates.

\section{Discussion}\label{sec:discussion}

As quantum computers become more available to the non-specialists, there is need for  tools that can be used and understood by users who are not interested in the inner workings of the machine. Benchmarking protocols have usually been designed with the experimental physicist in mind, often to quantify performance in terms of noise per gate, and usually with an outlook towards fault tolerance \cite{Lu2014a,Knill2008a}. The benchmarking experiments we used in this work were designed with the end user of a small noisy quantum processor in mind. Using two protocols based on the DQC1 trace estimation algorithm we benchmarked IBM 14- and 16-qubit processors and showed how performance degrades with circuit depth, in one case for the useful task of distinguishing knots.

In the first set of experiments we looked at how visibility drops as a function of circuit depth, and showed that coherent errors can be particularly harmful. For example we noted a spike in visibility (Fig.~\ref{fig:visplot}) for the DQC1 experiment with Qubit 11 on the  "Rüschlikon". We also observed an undesired buildup of an imaginary component (coherence in the $\sigma_y$ axis) as the gate count increased. This was evident as early as 10 CNOT gates in a 2-qubit experiment (Fig.~\ref{fig:res1}). Surprisingly, we noted that the performance is not reduced (and was even enhanced on average) in a 4-qubit experiment compared to the 2-qubit experiments when results with a similar CNOT gate count were compared (Fig.~\ref{fig:res1} and Fig.~\ref{fig:visplot}). Finally, we saw that the results were not consistent when the same experiment was performed at different times (Fig.~\ref{fig:diffdays}). However, the qualitative results regarding performance at different gate counts remained the same. 

In the second set of experiments we used 2 qubits on the IBM "Melbourne" to estimate the Jones polynomial at the 5th root of unity for various knots.  
While the results deviated from theory (Fig.~\ref{fig:JPdist}) it was possible to compare knots at low gate counts (Fig.~\ref{fig:JPresults}). 
However, at higher gate counts both depolarizing and systematic errors start to dominate the results. 
While depolarizing noise can be countered by repeating the experiment more times and normalizing \footnote{Repeating  experiments is generally not a scalable technique.}, the systematic errors (which are not constant in time) are difficult to deal with even in a small circuit.  We note that these errors prevented us from outperforming the 4-qubit liquid state NMR experiment \cite{Passante2009}.\footnote {However this is not an apples-to-apples comparison since (without fault tolerance) a 4 qubit machine is expected to outperform a 14 qubit machine in a $\le 4$ qubit experiment.} 

The relatively simple benchmarking procedure leads us to the conclusion that at least from the end user's perspective, a major issue with current small noisy quantum computers is  constantly changing environmental conditions that leads to frequently changing systematic errors.  While it is clear that this is a major engineering challenge for superconducting architectures due to the sensitivity to environmental conditions, it is worthwhile considering alternative architectures that may be more stable. A different approach might be a method to reduce systematic errors on the software side for example by using wider and shallower circuits or to turn these into statistical errors by using various randomization techniques. 



\acknowledgements 
We acknowledge use of the IBM Q for this work. The views expressed are those of the authors and do not reflect the official policy or position of IBM or the IBM Q team. This work was partially supported by the CIFAR “Quantum Information Science” programme and an NSERC grant “Experimental Quantum Information, Quantum Measurement, and Quantum Foundations With Entangled Photons and Ultracold Atoms” via Aephraim Steinberg’s research group.
KBF acknowledges the NSERC PDF programme for funding.

\bibliographystyle{plainnat}
\bibliography{bib}

\begin{thebibliography}{27}
\providecommand{\natexlab}[1]{#1}
\providecommand{\url}[1]{\texttt{#1}}
\expandafter\ifx\csname urlstyle\endcsname\relax
  \providecommand{\doi}[1]{doi: #1}\else
  \providecommand{\doi}{doi: \begingroup \urlstyle{rm}\Url}\fi

\bibitem[Cod()]{Code}
{The code used for running the algorithms is available online via GitHub}.
\newblock \url{https://github.com/agnostiQ/DQC1-knots}.

\bibitem[IBM({\natexlab{a}})]{IBM-m}
14-qubit backend: Ibm q team, “ibm q 14 melbourne backend specification
  v1.1.0,” (2018). retrieved from
  https://github.com/qiskit/ibmq-device-information/tree/master/backends/melbourne/,
  {\natexlab{a}}.

\bibitem[IBM({\natexlab{b}})]{IBM-r}
16-qubit backend: Ibm q team, “ibm q 16 rüschlikon backend specification
  v1.1.0,” (2018). retrieved from
  https://github.com/qiskit/ibmq-device-information/blob/master/backends/rueschlikon/,
  {\natexlab{b}}.

\bibitem[IBM({\natexlab{c}})]{IBMQ}
{IBM Quantum Experience}.
\newblock \url{https://www.research.ibm.com/ibm-q}, {\natexlab{c}}.

\bibitem[Aharonov et~al.(2009)Aharonov, Jones, and Landau]{Aharonov2009}
Dorit Aharonov, Vaughan Jones, and Zeph Landau.
\newblock {A polynomial quantum algorithm for approximating the jones
  polynomial}.
\newblock \emph{Algorithmica (New York)}, 2009.
\newblock ISSN 01784617.
\newblock \doi{10.1007/s00453-008-9168-0}.

\bibitem[Baez and Muniain(1994)]{baez1994gauge}
John Baez and Javier~P Muniain.
\newblock \emph{Gauge fields, knots and gravity}, volume~4.
\newblock World Scientific Publishing Company, 1994.

\bibitem[Boyer et~al.(2017)Boyer, Brodutch, and Mor]{Boyer2017}
Michel Boyer, Aharon Brodutch, and Tal Mor.
\newblock {Entanglement and deterministic quantum computing with one qubit}.
\newblock \emph{Physical Review A}, 95\penalty0 (2):\penalty0 022330, feb 2017.
\newblock ISSN 2469-9926.
\newblock \doi{10.1103/PhysRevA.95.022330}.
\newblock URL \url{https://link.aps.org/doi/10.1103/PhysRevA.95.022330}.

\bibitem[Emerson et~al.(2007)Emerson, Silva, Moussa, Ryan, Laforest, Baugh,
  Cory, and Laflamme]{Emerson2007a}
Joseph Emerson, Marcus Silva, Osama Moussa, Colm Ryan, Martin Laforest,
  Jonathan Baugh, David~G. Cory, and Raymond Laflamme.
\newblock {Symmetrized characterization of noisy quantum processes}.
\newblock \emph{Science}, 2007.
\newblock ISSN 00368075.
\newblock \doi{10.1126/science.1145699}.

\bibitem[Figgatt et~al.(2017)Figgatt, Maslov, Landsman, Linke, Debnath, and
  Monroe]{Figgatt2017}
C.~Figgatt, D.~Maslov, K.~A. Landsman, N.~M. Linke, S.~Debnath, and C.~Monroe.
\newblock {Complete 3-Qubit Grover search on a programmable quantum computer}.
\newblock \emph{Nature Communications}, 8\penalty0 (1):\penalty0 1918, dec
  2017.
\newblock ISSN 2041-1723.
\newblock \doi{10.1038/s41467-017-01904-7}.
\newblock URL \url{http://www.nature.com/articles/s41467-017-01904-7}.

\bibitem[Jones(1987)]{Jones1987}
V.~F.~R. Jones.
\newblock Hecke algebra representations of braid groups and link polynomials.
\newblock \emph{Annals of Mathematics}, 126\penalty0 (2):\penalty0 335--388,
  1987.
\newblock ISSN 0003486X.
\newblock \doi{10.2307/1971403}.

\bibitem[Jones(1985)]{Jones1985}
Vaughan~F.R. Jones.
\newblock {A polynomial invariant for knots via Von Neumann algebras}.
\newblock \emph{Bulletin of the American Mathematical Society}, 12\penalty0
  (1):\penalty0 103--111, 1985.
\newblock ISSN 02730979.
\newblock \doi{10.1090/S0273-0979-1985-15304-2}.

\bibitem[Knill and Laflamme(1998)]{Knill1998a}
E.~Knill and R.~Laflamme.
\newblock {Power of One Bit of Quantum Information}.
\newblock \emph{Physical Review Letters}, 81\penalty0 (25):\penalty0
  5672--5675, dec 1998.
\newblock ISSN 0031-9007.
\newblock \doi{10.1103/PhysRevLett.81.5672}.
\newblock URL \url{http://link.aps.org/doi/10.1103/PhysRevLett.81.5672}.

\bibitem[Knill et~al.(2008)Knill, Leibfried, Reichle, Britton, Blakestad, Jost,
  Langer, Ozeri, Seidelin, and Wineland]{Knill2008a}
E.~Knill, D.~Leibfried, R.~Reichle, J.~Britton, R.~B. Blakestad, J.~D. Jost,
  C.~Langer, R.~Ozeri, S.~Seidelin, and D.~J. Wineland.
\newblock {Randomized benchmarking of quantum gates}.
\newblock \emph{Physical Review A - Atomic, Molecular, and Optical Physics},
  2008.
\newblock ISSN 10502947.
\newblock \doi{10.1103/PhysRevA.77.012307}.

\bibitem[Lackenby(2015)]{Lackenby2015}
Marc Lackenby.
\newblock {A polynomial upper bound on Reidemeister moves}.
\newblock \emph{Annals of Mathematics}, 2015.
\newblock ISSN 0003486X.
\newblock \doi{10.4007/annals.2015.182.2.3}.

\bibitem[Lu et~al.(2014)Lu, Li, Trottier, Li, Brodutch, Krismanich, Ghavami,
  Dmitrienko, Long, Baugh, and Laflamme]{Lu2014a}
Dawei Lu, Hang Li, Denis-Alexandre Trottier, Jun Li, Aharon Brodutch, Anthony~P
  Krismanich, Ahmad Ghavami, Gary~I Dmitrienko, Guilu Long, Jonathan Baugh, and
  Raymond Laflamme.
\newblock {Experimental Estimation of Average Fidelity of a Clifford Gate on a
  7-qubit Quantum Processor}.
\newblock \emph{arXiv 1411.7993}, nov 2014.

\bibitem[Lu et~al.(2017)Lu, Li, Li, Katiyar, Park, Feng, Xin, Li, Long,
  Brodutch, Baugh, Zeng, and Laflamme]{Lu2017a}
Dawei Lu, Keren Li, Jun Li, Hemant Katiyar, Annie~Jihyun Park, Guanru Feng, Tao
  Xin, Hang Li, Guilu Long, Aharon Brodutch, Jonathan Baugh, Bei Zeng, and
  Raymond Laflamme.
\newblock {Enhancing quantum control by bootstrapping a quantum processor of 12
  qubits}.
\newblock \emph{npj Quantum Information}, 3\penalty0 (1):\penalty0 45, dec
  2017.
\newblock ISSN 2056-6387.
\newblock \doi{10.1038/s41534-017-0045-z}.
\newblock URL \url{http://www.nature.com/articles/s41534-017-0045-z}.

\bibitem[Morimae et~al.(2017)Morimae, Fujii, and Nishimura]{Morimae2017}
Tomoyuki Morimae, Keisuke Fujii, and Harumichi Nishimura.
\newblock {Power of one nonclean qubit}.
\newblock \emph{Physical Review A}, 2017.
\newblock ISSN 24699934.
\newblock \doi{10.1103/PhysRevA.95.042336}.

\bibitem[Park et~al.(2018)Park, Rhee, and Lee]{Park2018}
Daniel~K. Park, June Koo~K. Rhee, and Soonchil Lee.
\newblock {Noise-tolerant parity learning with one quantum bit}.
\newblock \emph{Physical Review A}, 2018.
\newblock ISSN 24699934.
\newblock \doi{10.1103/PhysRevA.97.032327}.

\bibitem[Passante et~al.(2009)Passante, Moussa, Ryan, and
  Laflamme]{Passante2009}
G.~Passante, O.~Moussa, C.~A. Ryan, and R.~Laflamme.
\newblock {Experimental Approximation of the Jones Polynomial with One Quantum
  Bit}.
\newblock \emph{Physical Review Letters}, 103\penalty0 (25):\penalty0 250501,
  dec 2009.
\newblock ISSN 0031-9007.
\newblock \doi{10.1103/PhysRevLett.103.250501}.
\newblock URL \url{http://link.aps.org/doi/10.1103/PhysRevLett.103.250501}.

\bibitem[Passante(2012)]{Passante2012a}
Gina Passante.
\newblock \emph{{On Experimental Deterministic Quantum Computation with One
  Quantum Bit ( DQC1 )}}.
\newblock PhD thesis, University of Waterloo, 2012.

\bibitem[Pokharel et~al.(2018)Pokharel, Anand, Fortman, and
  Lidar]{Pokharel2018}
Bibek Pokharel, Namit Anand, Benjamin Fortman, and Daniel Lidar.
\newblock {Demonstration of fidelity improvement using dynamical decoupling
  with superconducting qubits}.
\newblock jul 2018.
\newblock \doi{10.1103/PhysRevLett.121.220502}.
\newblock URL \url{http://arxiv.org/abs/1807.08768
  http://dx.doi.org/10.1103/PhysRevLett.121.220502}.

\bibitem[Preskill(2018)]{Preskill2018quantumcomputingin}
John Preskill.
\newblock Quantum {C}omputing in the {NISQ} era and beyond.
\newblock \emph{{Quantum}}, 2:\penalty0 79, August 2018.
\newblock ISSN 2521-327X.
\newblock \doi{10.22331/q-2018-08-06-79}.
\newblock URL \url{https://doi.org/10.22331/q-2018-08-06-79}.

\bibitem[Pullin(1994)]{pullin1994}
Jorge Pullin.
\newblock Knot theory and quantum gravity in loop space: A primer.
\newblock \emph{AIP Conference Proceedings}, 317\penalty0 (1):\penalty0
  141--190, 1994.
\newblock \doi{10.1063/1.46852}.
\newblock URL \url{https://aip.scitation.org/doi/abs/10.1063/1.46852}.

\bibitem[Rudolph(2016)]{Rudolph16}
Terry Rudolph.
\newblock {Why I am optimistic about the silicon-photonic route to quantum
  computing}.
\newblock jul 2016.
\newblock URL \url{http://arxiv.org/abs/1607.08535}.

\bibitem[Shor and Jordan(2007)]{Shor2007}
Peter~W Shor and Stephen~P Jordan.
\newblock {Estimating Jones polynomials is a complete problem for one clean
  qubit}.
\newblock \emph{Quantum Information and Computation Vol.}, 8:\penalty0 pg.681,
  jul 2007.

\bibitem[Takita et~al.(2017)Takita, Cross, C{\'{o}}rcoles, Chow, and
  Gambetta]{Takita2017}
Maika Takita, Andrew~W Cross, A~D C{\'{o}}rcoles, Jerry~M Chow, and Jay~M
  Gambetta.
\newblock {Experimental Demonstration of Fault-Tolerant State Preparation with
  Superconducting Qubits}.
\newblock 2017.
\newblock \doi{10.1103/PhysRevLett.119.180501}.
\newblock URL
  \url{https://journals.aps.org/prl/pdf/10.1103/PhysRevLett.119.180501}.

\bibitem[Wallman et~al.(2015)Wallman, Granade, Harper, and
  Flammia]{Wallman2015}
Joel Wallman, Chris Granade, Robin Harper, and Steven~T. Flammia.
\newblock {Estimating the coherence of noise}.
\newblock \emph{New Journal of Physics}, 2015.
\newblock ISSN 13672630.
\newblock \doi{10.1088/1367-2630/17/11/113020}.

\end{thebibliography}

\onecolumn\newpage
\appendix

\end{document}